# AGN in Gaia Alerts: from flares to Changing Look Quasars

M. Dennefeld [1], T. Pursimo [2], C. Carvalho [1]

E. Biancalani[3*], M. A. Díaz Teodori[2*], O. Dürfeldt Pedros[4*], M. A. Fetzner Keniger[5*], A. Kasikov[6*], N. Koivisto[7*], J. Martikainen[8*], K. Matilainen[7*], J. Sinkbak Thomsen[9*], J. Terwel[10*], and A. Viitanen [11*]

1 Sorbonne University & CNRS, Institut d'Astrophysique de Paris (IAP), F-75014 Paris
2 Nordic Optical Telescope (NOT), E-38700 Santa Cruz de La Palma
3 University of Maryland, College Park, USA
4 Technical University of Denmark, DTU Space
5 University of Warwick, Dept. of Physics, UK
6 University of Tartu, Tartu Observatory, Estonia
7 University of Turku, Dept. of Physics and Astronomy, Finland
8 CSIC, Inst. de Astrofisica de Andalucia, Granada, Spain
9 Universita di Bologna, Dipt. di Fisica e Astronomia, Italy
10 Trinity College Dublin, School of Physics, Dublin, Ireland
11 INAF-Osservatorio Astronomico di Roma, Italy

\* Student Support Astronomer at the Nordic Optical Telescope

## Abstract

The Gaia Alerts system is providing alerts on a variety of objects displaying a significant photometric change detected by the Gaia satellite from one passage to the next one over the same region of the sky. Among the ~23000 alerts published until the end of 2022, around 13% concern AGN or quasar candidates. At the same time, a different method to detect variations specifically in galactic nuclei was tested on Gaia data during a one year period only, leading to another set of candidates. We have embarked on a spectroscopic ground-based follow-up of some of those, to establish their true nature, and reveal the various phenomena leading to a change in magnitude of those AGN. The present paper deals with radio-quiet objects, while the radio-loud ones will be described in a companion paper. We confirm, on one hand, the AGN/quasar nature of 64 new candidates alerted by Gaia, and, on the other hand, obtained second-epoch spectra of over 200 known AGN also alerted for large photometric variations. The observed phenomena show a large variety described in this paper: from Flares to Tidal Disruption Events (TDEs) and a large number of Changing Look Quasars (CLQs, 56 secure ones, plus 23 probable ones), not forgetting some rarer events like SNe, microlensing events or Extreme Coronal Line Emitters. This sample shows that variability is an excellent tool to detect new quasars, especially radio-quiet ones that otherwise would be undetected, and that a significant fraction of variable AGN/quasars, around 10%, presents the CLQ phenomenon. Some of the new CLQ's are followed-up to monitor further changes and measure time scales.

# I) Introduction

Since the identification of the first Seyfert galaxies (Seyfert, 1943) and the discovery of the first quasar, 3C 273 (Schmidt, 1963), various techniques were employed to find new candidates and the zoo of extragalactic objects increased accordingly. In the optical, objective prism and other surveys revealed objects with compact cores and/or blue excess (Markarian, 1963; Haro &Luyten, 1960; Zwicky, 1964; and subsequent papers in the series), while the search of optical counterparts to radio sources lead to the identification of radio galaxies and radio-loud quasars. It took a while, however, until some order could be put in this diversity. First, it is now established that Seyfert galaxies and quasars are of the same nature, the only difference being in their distance and luminosity (with an arbitrary separation put at $M_V = -23$ (Véron-Cetty & Véron, 1984) so that the surrounding galaxy was detected or not. Second, more importantly, in proposing the unified scheme (Antonucci, 1993): a central black-hole accretes matter via an accretion disk, surrounded by a dusty torus, a "Broad Line" region (BLR) and a "narrow-line" region (NLR) (Urry & Padovani, 1995). If the line of sight from the observer sees directly the central engine and the BLR, one has a Seyfert 1 galaxy (or type 1 quasar), while if the view is obscured by the torus, one has a Sey 2 or a type 2 quasar. The radio properties are an additional feature, with a central emission, plus eventually a jet coming out from the center (either directed towards the observer, or not), which can excite radio-lobes at larger distances. The objects are qualified as "radio-quiet" if the flux of the radio emission is less than ten percent of the optical emission (e.g. 5GHz radio over B band optical, all expressed in Janskys, Kellermann et al. 1989) but this is again an arbitrary limit initially set by the sensitivity limit of the first radio telescopes.

In this scheme, things are stable over time, as the viewing angle from the observer does not usually change significantly over a human life time (excluding the radio jet). Some small luminosity changes are however possible but are limited to a few tenths of magnitude: it has been shown, for instance, that optical variations at the 0.03 magnitude level are seen in at least 90% of quasars in a well-studied sample like the SDSS Stripe 82 (Sesar et al. 2007). Large variations in radio flux, like in blazars, are not discussed here. Larger, optical luminosity changes have been used to refine the knowledge of the internal structure of AGN: the observed time delays between changes at various wavelengths (the reverberation mapping technique, Peterson et al., 2004) are consistent with the unified scheme. However, things are changing since the advent of large all-sky surveys, starting with the Catalina Real Time Survey (CRTS, Drake et al. 2009) and the Sloan Digital Sky Survey (SDSS, Ahumada et al. 2020 for the DR16 used here): by covering a given region of the sky several times over the years, they allowed to notice large magnitude changes in some quasars. The Zwicky Transient Facility (ZTF, Bellm et al. 2019) follows a given region of the sky every three nights and allows to discover shorter variations in AGN's and Quasars. Other surveys, although dedicated mainly to other targets, in particular Near Earth Asteroids like ATLAS (Smith et al. 2020) or Supernovae (like ASAS-SN, Kochanek et al. 2017), contribute also to the study of AGN variability. Various phenomena are observed: it can be a "flare", a short time change as sometimes also seen in blazars, possibly due to changes in the accretion rate, or instabilities in the local environment of the central AGN. It can be a Tidal Disruption Event (TDE), when a star or other object is reaching too close to the black hole and gets disrupted by gravitational forces: several TDE's have been discovered recently and a classification has been proposed by van Velzen et al. (2021). This is similar to what is seen in our own Galaxy ( an in-spiraling star near Sgr A, Genzel et al. 2012). Or it can be a change in obscuration between the observer and the central engine, or another unexplored mechanism.

Of particular interest are the so-called Changing Look Quasars, CLQ's: by targeting quasars with changes larger than one magnitude in the SDSS survey, MacLeod et al. (2016) found that the luminosity change is accompanied by a change in spectrum.  The broad lines appear when the luminosity increases, and disappear when it decreases, that is a change from Sey 2 to Sey 1 and vice-versa, and this, over a time scale of a few years at most. A few such cases were known before, discovered by chance, like in NGC 4151 (Penston & Perez, 1984) or NGC 7582 (Aretxaga et al. 1999), but these few could be ascribed to peculiar cases. But the actual multiplication of such cases makes it difficult to reconcile with the standard unified model. In fact, even more cases have been discovered since then, for instance by Graham et al. (2020) searching in the CRTS. In the SDSS itself, Yang et al. (2018) and Potts & Villforth (2021) found more cases by comparing directly second epoch spectra to first ones, irrespective of magnitude changes. We have ourselves found some more by comparing first and second epoch SDSS and LAMOST spectra for quasars alerted by Gaia for a large magnitude change (> 0.5 mag, see below; Huo et al. 2020). It seems thus clear that, as seen from those surveys, quasars and Seyferts can exhibit large magnitude changes (up or down) more frequently than previously thought.  Furthermore, when the object is previously unknown and not associated to a radio-source, such large magnitude changes are a good indicator for a candidate quasar, which can then only be confirmed by spectroscopy. We have thus searched the whole Gaia-Alerts catalogue for objects either associated to known AGN (Seyferts or quasars) or flagged as candidate quasars and obtained second epoch spectroscopy for some of them, revealing the variety of cases detected by Gaia. In Section 1, we describe the Gaia-Alerts and the selection of objects, with its limitations. In Section 2, we detail the new quasars confirmed by our spectroscopy. In Section 3 we describe the variety of cases found in the Gaia Alerts, with emphasis on CLQ's and TDE's, and discuss the possible interpretations. We summarise and conclude in the last section.

## II)   The GAIA Alerts

Launched in December 2013, the ESA/Gaia satellite aims at producing, as a core product, accurate and precise astrometry and photometry for over 1.8 billion celestial sources (Gaia Collaboration, 2016). The data are released periodically, in successive installments, the latest of which was released in June 2022 (Gaia DR3, Gaia collaboration 2022). The satellite is spinning around an axis inclined at 45° from the Sun, one rotation taking 6 hours. With its two telescopes separated by 106° along the circle, the satellite is pointing regularly at the same area of the sky, the shortest repetition being of 106 minutes, then 6 hours.  But, because of the precession of the satellite's axis, the next visit can be soon (6h for an object in the center of the field) or later (for objects closer to the edge of the field of view, which fall out of view at the next rotation), on average about 63 days, depending on the ecliptic latitude. These regular visits, expected to be of several hundred over the lifetime of the satellite (which ceased operations on January 15$^{th}$, 2025), do allow the detection of transients or variable objects, but would not be fully exploitable with Data Releases separated by several years. For this reason, an Alert system has been implemented, to complement the main mission, and is operated in Cambridge (England). A full description of this Alert system has been given by Hodgkin et al. (2021): we recall here only the main aspects required for our analysis of AGN. The Alerts are based on photometric variability, based on Gaia's own photometry alone (without reference to previous, ground-based photometry).  For an Alert to be issued, the object should display a delta-magnitude in the Gaia G band, initially set to 1 mag, at least 3 σ above the baseline flux, or, for sources with smoother behavior, 0.15 mag but at least 6 σ above the baseline flux. The thresholds for Alerts were later relaxed to 0.5 then 0.3 mag, from

2019 onwards, once the system's behavior was better understood.  To minimize the number of Alerts, regular variables are excluded and treated elsewhere (Gaia Collaboration et al. 2019). After excluding various artefacts (see Hodgkin et al. 2021), the remaining cases are all checked by a human scrutator before publication is decided, with indication of the possible nature of the Alert or association with existing catalogues (stars, galaxies, quasars, etc...). While the number of Alerts was few at the beginning, the rate increased progressively with experience gained, to reach at the end around a dozen per day. They also come now with a better characterization, with the help of the (uncalibrated) BluePrism (BP) and RedPrism (RP) very low-resolution spectra on board and association with external catalogues.

We work here with the 21713 Alerts published until 22d of October 2022, when a long (but temporary) stop in Alerts was noticed, due to upgrades in the hardware (but our follow-up observations are continuing with more Alerts, which reached 27076 in total at the end of operations of the satellite in January 2025). We have selected for possible follow-up spectroscopy, all those marked as associated to known Seyferts, AGN or quasars and those marked as candidates QSO's or quasars.  The variety of denominations in the comments (QSO, quasar, Seyfert, AGN, etc...) reflects the diversity of the scrutators and shows that the selection can by no means be considered as complete or homogeneous, thus making it difficult to perform good statistical studies. We have a total of 84 Alerts in known Seyfert galaxies (of either type, 1 or 2), 133 in AGN (among which 57 are marked candidate AGN), and   1336 in quasars (or QSO's), among which 367 marked as "candidate QSO's.  We have added also some cases of objects marked as associated with a GALEX, WISE or X-ray source (without mention of AGN or QSO), to fill in the observing plan, as such a detection could be a good indication of a non-thermal source. As spectroscopy was not always possible immediately after the Alert (for instance, the right ascension not being observable at the time), and because the Alert itself often arrives long after the start of the rise in flux (especially for slow but continuous rises, until the threshold for Alert of 0.5 or later 0.3 magnitudes was reached), the total light curve sampled by Gaia is an important tool also to assess the nature of the Alert.  Particularly for Seyfert 2's, the light-curve is often compatible with a Supernova, some of them confirmed spectroscopically, and these SNe are not considered here. We instead concentrate on the diversity of AGN and Quasars.

An alternative selection algorithm to better detect variations specifically in the center of galaxies was tested by Kostrzewa-Rutkowska et al. (2018) (herewith called the KR sample). They detected 482 nuclear transients over a one year period only (mid-2016 to mid-2017):  among them, 339 are marked unknown and only 4 were included in the routine Gaia Alerts. This method, although more efficient, could however not be implemented in the standard Alerts stream due to practical limitations. We have nevertheless included some of their objects in our follow-up program as observing time permitted: either known AGN to detect possible spectral variations, or some of the 339 unknown which, although catalogued in the SDSS photometric survey, had no available spectrum and are possible quasar candidates. As their light-curves have not been followed further after this one-year test-period, it is more difficult to assess the origin of their variations if no spectrum is available and they are not included in the overall statistics below.  They are discussed separately in the Appendix.

## II) Candidate quasars

We have first looked at the candidate quasars (or AGN), which amount to 424 in this sample. These candidates were selected by the Alert Team on the basis of a large photometric variability (the required amplitude of which has been lowered with time) and also on the shape of the (uncalibrated) low resolution BP and RP spectra, following simulations made by Blagorodnova et al. (2014) with various SED's, but we have no further details on the selection process. As not all the 424 candidates could be observed, we have also cross-correlated the Gaia-Alerts catalogue with the Swift X-ray telescope point source catalogue (Evans et al. 2020), to increase the probability to have bona-fide quasar candidates, and found 157 known AGN in common, plus 37 QSO candidates. We have then given priority to those candidates with X-rays data when establishing the observing list of candidates, thus introducing a positive bias in the selection. But as the main goal of this work is to find new CLQ's, a large part of the observing was devoted to objects showing large photometric variations and having already a reference spectrum somewhere, mainly in the SDSS (DR16 release, Ahumada et al. 2020).

Most of the observations were done with the Nordic Optical Telescope (NOT) in LaPalma, using the ALFOSC spectro-imager. with grating #4. Spectra cover a useful wavelength range from 380nm to 900nm, at a resolution of ~300 with a 1.3" entrance slit projecting as 3 pixels on the detector (in a few cases, a 1.0" slit was used for a resolution of ~400). Data reduction was done in the standard way, with bias correction, flat fielding, wavelength calibration and response curve determination through observations of standard stars. A calibration arc was usually obtained at the same position of the telescope, immediately before or after the science observation, to minimize redshift errors. As many of the observations were done as filler programs, we could usually not observe hot stars close enough to the objects to correct for telluric absorption lines, but those lines are marked by a circled + sign on the spectra. Some objects were also observed more recently with the new Mistral spectro-imager at the 1.93m telescope at Haute-Provence Observatory (OHP; Schmitt et al. 2024) from 2022 onwards, with a shorter spectral range (410-810 nm) and a higher spectral resolution of ~700. We also searched the literature for existing spectra, leading to a few more classifications, particularly for southern objects, e.g. with the 6dF survey (Jones et al. 2009).

In total, 33 of the 424 GAIA-Alerts quasar candidates could be observed, and also 33 of the 339 "unknowns" of the KR sample, which represents only a small ~ 10% sub-sample of the available candidates. The list of objects observed is given in **Table 1**, together with the lines detected and the measured redshift. Lines were fitted with Gaussian profiles to derive their position. Some illustrative spectra are shown in **Fig.1**. In a few cases, the spectrum is inconclusive, because of insufficient signal to noise ratio, either due to a too short exposure time (the objects were mostly observed as fillers during other programs or during twilight) or to poor weather conditions, and these will be re-observed at a later opportunity. In only rare cases was the obtained classification different from a quasar (e.g. Gaia 16bpi, but this one was selected at the very beginning of the program, when the Gaia Alerts selection method was not yet fully stabilised): this shows the power of variability as a tool to find new quasars. Once a redshift is available, it is possible to derive an absolute magnitude for the object (we adopt a flat Lambda cold dark matter cosmology with a Hubble-Lemaitre constant of 68 km/s/Mpc for this purpose), but we did not use it to distinguish between Seyfert galaxies and quasars, generally keeping the original Gaia classification: the usual limit of -23 is arbitrary, and the distinction should merely be based on whether or not one can distinguish the underlying galaxy, which cannot be done with Gaia data alone, as the images are not available. We end-up in confirming the AGN nature of 42 candidates (plus 22 from the KR sample), all given in **Table 1**.

Towards the end of this work appeared an analysis of the low-resolution BP-RP spectra (R~~50) of known or candidate QSO's in the Gaia DR3 release (Storey-Fisher et al. 2024 : Quaia, the Gaia-UnWise quasar catalog); they list a total of 755850 objects down to the Gaia-g magnitude of 20. We have added their results to our Table 1, which shows that, with a few exceptions only (marked by an * in Table 1), their redshift evaluation is fairly consistent with our spectroscopic measurement, although obviously with a much poorer accuracy than our results, due to the very low spectral resolution of the BP-RP spectra (we have truncated their z values, in our Table 1, to the first four decimals, as further decimals make no sense in view of their low resolution). A few of our objects are not in Quaia, possibly because they were detected as extended by Gaia.

## III) The variety of AGN phenomena in the Gaia Alerts

Besides candidate quasars discussed above, the Gaia Alerts in known AGN refer in fact to a variety of phenomena, which can in practice only be precisely characterised either by an optical spectrum obtained rapidly from the ground, or, if this was not possible, eventually by the full light-curve available only long after the Alert. From the variety of light-curves, we can distinguish several types of events: short duration flares, which can also include some supernovae; a smooth rise in luminosity followed by a symmetric decline or the reverse (where the decline is preceding the symmetric re-brightening); Tidal Disruption Events (TDE's) where a rapid increase in luminosity is followed by a shallower, power-law decline; smooth, long-term increase (or decrease) in luminosity which is often concomitant to significant spectral changes, the so-called Changing Look Quasars (CLQ's); or even a combination of more than one such cases. We will discuss all those in the next sections in turn.

### 1. Tidal Disruption Events (TDEs) and Coronal Line Emitters (CLEs)

TDEs refer to the disruption of a star passing too close to a Black Hole (BH) who rips it apart and swallows (part of) its material (see Gezari, 2021 for a review). Under simplified assumptions, the fallback rate of the debris, and thus the declining light-curve after the peak, should follow a form of $t^{-5/3}$; while the rise to peak is much faster, of the order of a few days. One TDE candidate from Gaia Alerts, Gaia19bpt, has been confirmed spectroscopically by van Velzen et al. (2021) and shown to be of the Hydrogen (H) type. Indeed, they identify three spectroscopic classes of TDEs: the H class where only hydrogen lines are seen in the spectrum; the rarer He class, where only He lines are seen; and the mixed H+He class which also shows Bowen fluorescence lines of OIII and NIII, transitions excited by the excitation and recombination of HeII. Five other TDE candidates (Gaia 18dpo, 19bvo,19bti, 19dhd and 19eks) discovered by other surveys, were also recovered by Gaia and their spectra obtained by van Velzen et al. (2021). We use these typical light-curves from Gaia to identify other possible TDE candidates in Gaia Alerts. We show, as examples, the light-curves of Gaia 19bpt and Gaia 20alj in **Fig. 2a**. Gaia20eub was also identified as a TDE, following its initial detection in the X-rays (Homan et al. 2023) and is shown in **Fig. 2b**.

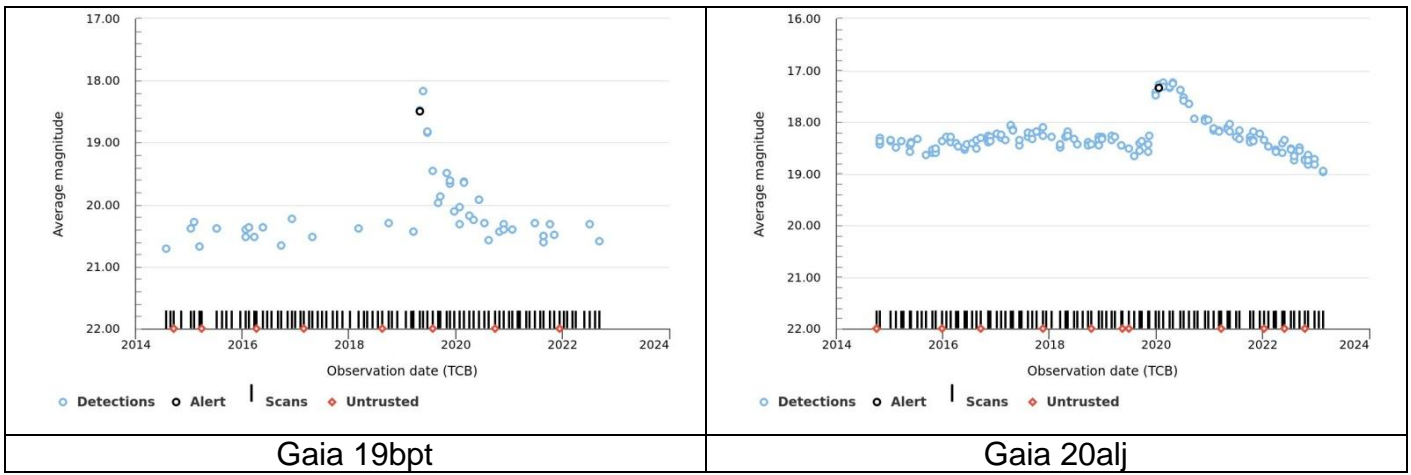

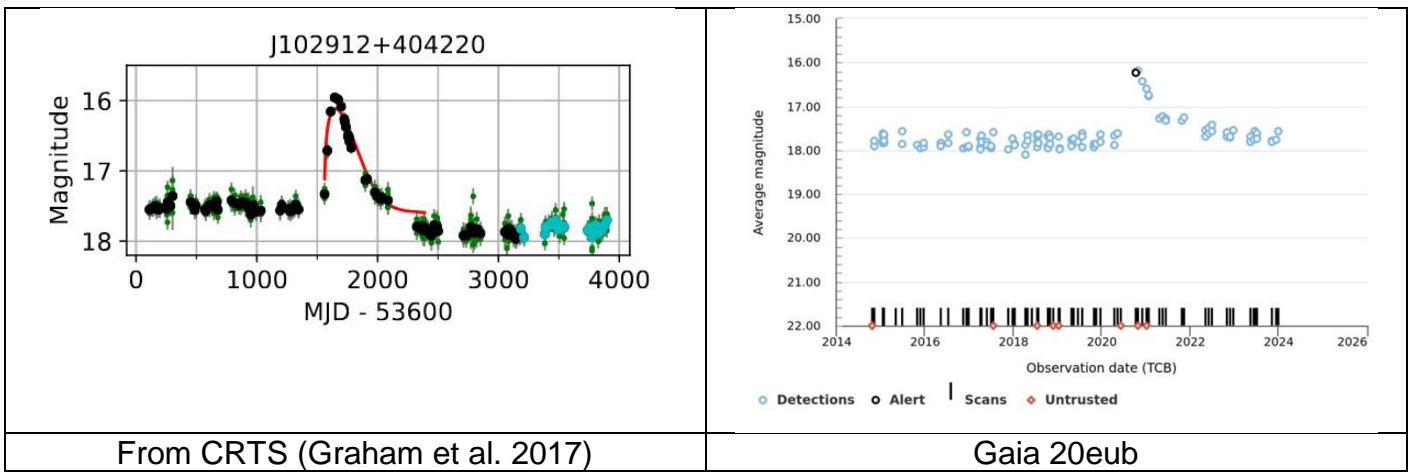

Fig. 2 a (top), b (middle): Gaia light-curves, Gaia g magnitude versus observing dates (in years, blue points; individual observations are marked by a black tick in abscissa). The date of the Alert is marked by a black point, instead of a blue one. The light curve of J102912+404220 is extracted from Graham et al. (2017)

The best example is Gaia 20alj, alerted on Jan. 22, 2020, when it reached a g-mag of 17.3, close to the peak. Its light-curve is very similar to those mentioned previously. Our first spectrum was obtained on 15th of February and showed some Balmer lines and the Bowen fluorescence complex of HeII 4686 + NIII 4640 (see **Fig. 3**), therefore suggesting a TDE of the H+He class. But this spectrum showed also several coronal lines of Fe, namely [FeX] 6374, [FeXI] 7892 and [FeXIV] 5304 Å, therefore putting the object in the rarer class of Extreme Coronal Line Emitters. We also identify a [NiXV] line at 8024 Å and possibly [FeVI] at 5632 Å. A series of weak blended lines around 8460 rest wavelength (observed 9195) could be from [FeI] multiplet 33F but would then imply that these lines are emitted far from (and shielded from) the central ionising source, which remains to be confirmed. In later spectra obtained on July 13th, 2021 and Feb.13th 2022 (**Fig. 3a**) the Bowen fluorescence lines have disappeared, while the coronal lines are still present. This is consistent with the suggestion that these coronal lines are produced by gas located further away from the nucleus (Wang et al. 2012). Four other examples with coronal lines were found in the Gaia Alert stream: Gaia 20cvt, 20fku, 21fiy and 22djl, all showing strong [FeVII] lines and their spectra are shown in **Fig. 3b**

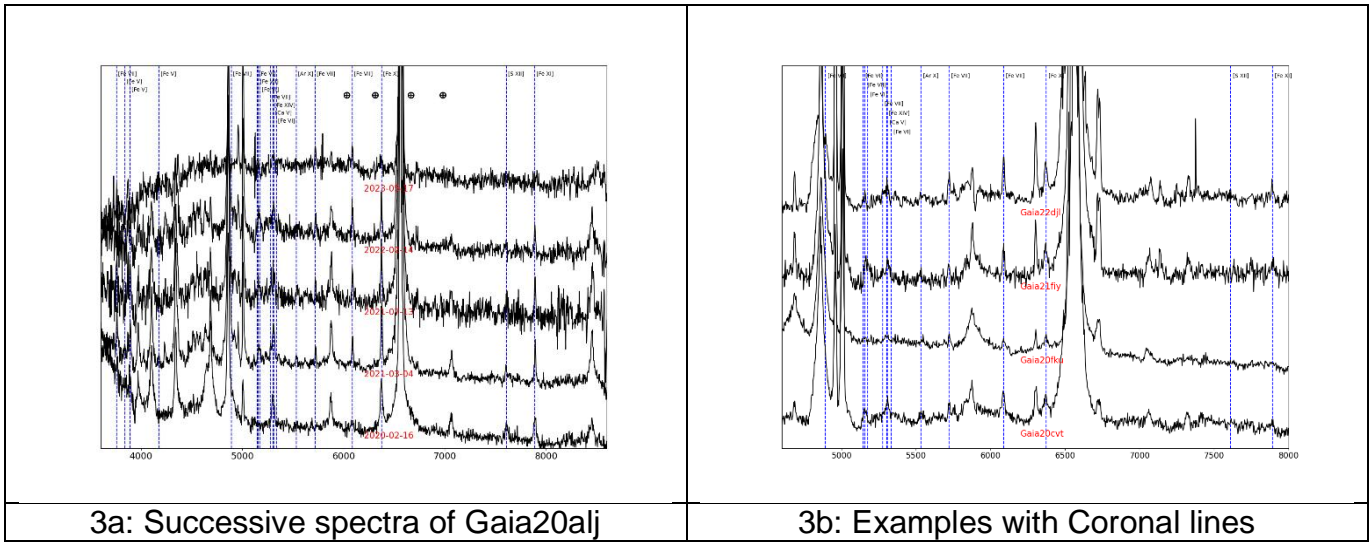

Fig. 3 a: Successive spectra of Gaia20alj, flux versus wavelength. The date of the spectrum is indicated in red, below it (from bottom to top: 02/2020, 03/2021, 07/2021, 02/2022, 05/2023, with a vertical offset). Vertical, blue dotted lines indicate the position of expected emission lines. Telluric absorption lines are marked with a circled + sign

Fig. 3 b From bottom to top, spectra of objects with coronal lines: Gaia 20cvt, 20fku, 21fiy and 22djl.

By contrast, Gaia 17cff has a similar light-curve shape **(Fig. 3c),** but, surprisingly, much lower scatter in the light curve, contrary to the previous objects. It was initially classified as a SN-II (Xhakaj et al. 2017) but later reassigned to an AGN in view of its absolute magnitude of  - 23 reached at peak. The published spectrum (obtained shortly after peak, Arcavi et al., 2017) shows moderately broad Balmer lines which could indicate an H-type TDE, or a Narrow Line Seyfert 1 (NLS1) galaxy.  We obtained a late-time spectrum (April 2023, Fig. 3c) of the host which does clearly show a NLS1 (FWHM 1100 km/s, identical within the uncertainties to the width in the earlier spectrum) with significant FeII emission and a slightly less blue continuum. While the decreasing part of its light-curve is compatible with the canonical -5/3 slope expected for TDEs, we note a second, lower peak in Feb. 2020, but not well sampled by the scarce Gaia coverage, raising the question of a recurring TDE or a binary system. The available spectra do not help to properly classify the 17cff event, but confirm that variations are a good way to reveal otherwise unconspicuous AGN's, and that NLS1's are more likely to reveal TDE's, due to their lower BH mass, and hence smaller horizon.  We note that Gaia17cff has a light curve quite similar to J102912+404220 in Graham et al. (2017, see **Fig. 2b**), which is quite unique in their sample, but was not further discussed there: the latter has an SDSS spectrum showing a NLS1 type (with FeII emission), thus comforting our interpretation as a TDE.

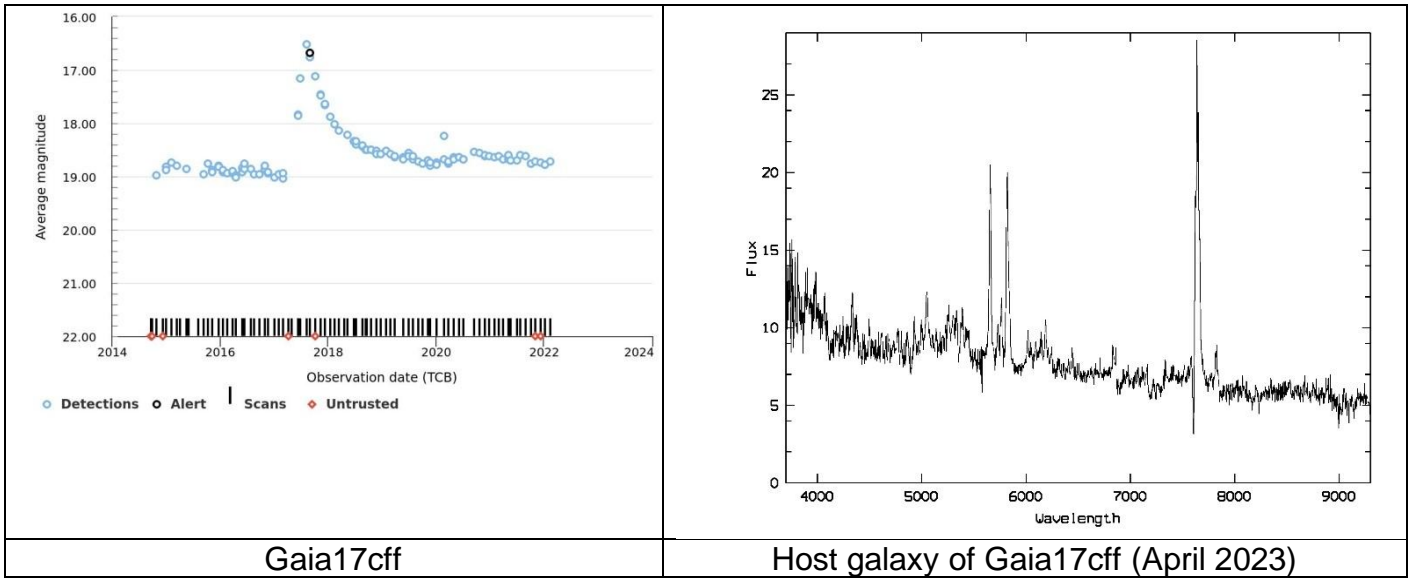

| Gaia17cff | Host galaxy of Gaia17cff (April 2023) |

Fig. 3c  Gaia17cff: Light-curve (left, g magnitude versus observing date), classification spectrum (right,

flux (units 10E-16 ergs/cm$^2$/sec/Å versus wavelength)

Gaia21aid is another example. Alerted on Jan. 18$^{th}$ 2021, it is a WISE and SDSS galaxy, but had no available redshift. We obtained 3 spectra at different epochs, as shown in **Fig. 3d**. The first spectrum obtained shortly before the maximum displays a weak, broad line component and the standard HeII+NIII complex characteristic of a TDE. This complex disappears with time, while the broad lines strengthen and an FeII complex appears, suggesting the host galaxy is a NLS1, a favorable situation to reveal TDE's, as discussed before for Gaia17cff.

For Gaia19axp, we obtained a spectrum shortly after the Alert, which proved to be still before the maximum, and this spectrum shows the characteristic NIII + HeII complex **(Fig.3d)**, classifying this case as a TDE event also, confirmed by the later light curve.

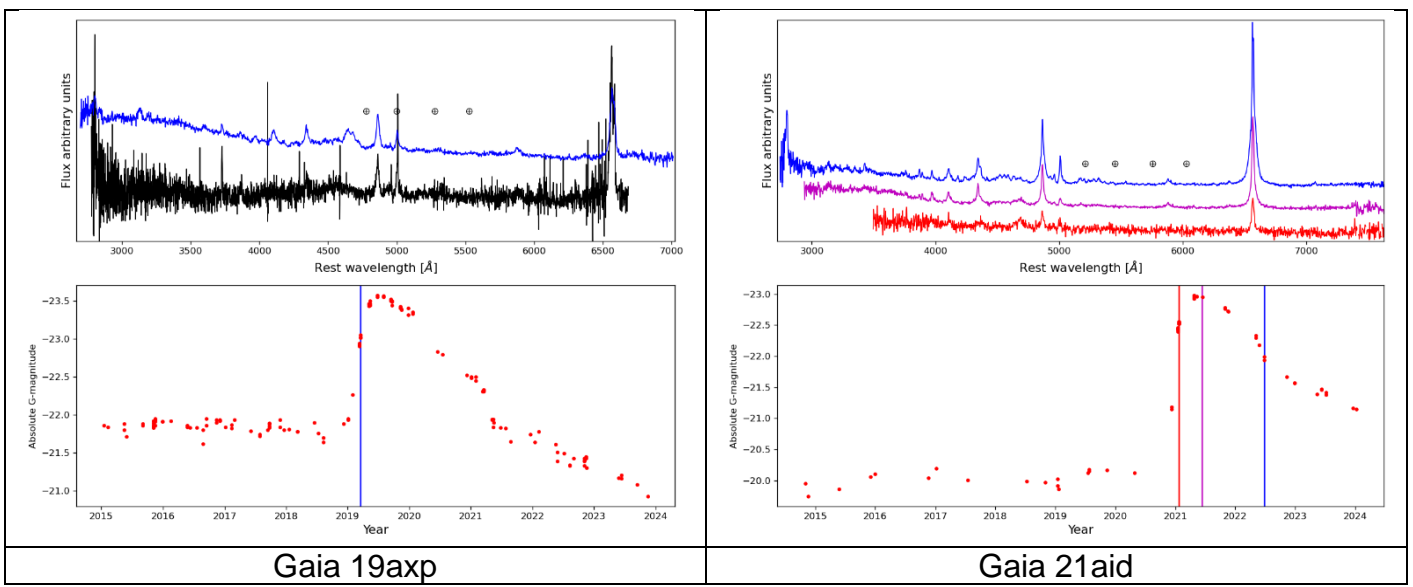

| Gaia 19axp | Gaia 21aid |

Fig. 3d  Gaia19axp (left) and Gaia21aid (right)

The light curve (Gaia absolute magnitude versus time) is at the bottom, vertical lines indicate the dates of the spectra obtained. The Gaia g magnitude has been transformed into absolute G magnitude with the help of our redshift, using $H_0$ = 68 km/s/Mpc. The spectra themselves are displayed in the upper part (flux versus wavelength; telluric lines are marked with a circled +), with a color corresponding to the one of the vertical line in the light curve. The SDSS reference spectrum is the lower one, in black (only available for Gaia19axp).

## 2. Changing Look Quasars

When the Alert is associated to a known AGN or quasar, this means that a classification spectrum exists somewhere in the databases. The Alert indicates the amplitude and timescale of the detected photometric change, and a new spectrum should then help to detect possible spectroscopic changes also. Among the hundreds of AGN alerted this way, we could obtain a second epoch spectrum for 99 of them, plus 15 from the KR sample. As the alert itself is sometimes issued long after the beginning of the change (particularly in the case of a smooth increase, until the change reaches the fixed threshold of 0.5 or 0.3 magnitudes), and then the object is not necessarily immediately observable from the ground, this second epoch spectrum is sometimes obtained long after the beginning of the event. It can then not be used directly to measure timescales or delays between the photometric and the spectroscopic changes. We have detected, by visual inspection of the spectra, 49 cases where obvious changes in the BLR and in the blue continuum are apparent in the new spectrum with respect to the archival one, plus 17 cases where some smaller changes may be present but need to be measured more precisely. In the KR sample, 7 changes are obvious, while 6 others need to be better estimated.

The list of objects observed spectroscopically, together with their properties, is presented in **Table 2.** The light curves (with marked dates of the spectrum) and spectra for some illustrative cases are shown in **Fig.4** (all the spectra will be available on-line). While a detailed, quantitative analysis is deferred to a forthcoming paper, the general behavior can be described as follows: an increase in the broad lines goes along with an increase in the blue continuum, with sometimes also a smaller increase of the MgII lines (when available in the observed spectral range). This goes along with an apparent decrease of the narrow lines ([OIII], or [NII]) intensities, but needs to be measured precisely. It is more difficult for [NII] as those lines are often covered when Halpha is broadening. Before trying to interpret those changes, some caveats are needed. First, the SDSS reference spectra were taken with a 3" entrance fiber, while we used a 1.3", or sometimes a 1.0" slit, so the apparent changes in the NLR with respect to the continuum may be partly due to this sampling difference. Second, the dominance of cases with increased BLR and blue continuum is a selection effect, as we observed preferentially objects with brighter magnitudes, in view of the limits with a 2m class telescope. We see a variety of initial cases leading to the appearance or increase of the broad line components, including initial Liners (e.g. Gaia20ctd), Sey 2, Sey 1.9 and some Sey1 where the BLR displays a strong increase. If we take the numbers at face value, we have 59 secure, confirmed cases of changing look, and up to 91 if we include less secure cases (which need to be measured), out of a sample of 200 candidates alerted for variations (excluding the blazars), that is roughly a 50% success rate, assuming there is no particular bias in the initial selection by Gaia Alerts. The phenomenon of "Changing Look" seems therefore to be much more common than initially thought, and its interpretation will be key for the physics of the innermost regions of AGN. Graham et al. (2020) have identified 111 cases of spectroscopic changes following a photometric change detected by the CRTS, but we have only one object in common, Gaia 20bsq. Similarly, we have only one object in common in our main sample, Gaia 20axp, with the sample of 61 of Green et al. (2022) (and two more in the later extension of our work after 2022). Clearly, selection effects in the initial samples are important.

Ricci & Trakhtenbrot (2023) distinguish between two main possible scenarii: changing obscuration (CO-AGN); or changing state (CS-AGN, i.e. a change in the central luminosity). At this stage, we do not have enough data to progress in the interpretation for our objects. A change in obscuration is possible, but a measure of the H$\alpha$/H$\beta$ ratio, even if changing, is of little help as there is no canonical value of the value it should have in the absence of obscuration (contrary to classical HII regions, e.g. their Case B). We need a measure of the gas and dust column densities to estimate the obscuration (hence X-rays and IR observations). Changes in the accretion rate or changes in the structure of the accretion disk are another possibility. But it is difficult to conclude with the presently available data; we need to be able to measure time delays between the changes in the continuum (central engine) and the changes in the BLR, that is the propagation time from the center to the BLR. The present scarce sampling by Gaia, and the delayed spectroscopic follow-up has not allowed to measure this until now. We need also multi-wavelength data: in the optical, to measure delays between the photometric and the spectroscopic changes; in the IR and in the X-rays, to measure changes in obscuration and in gas column-densities. Spectro-polarimetry would also be a valuable tool, but such instruments are rarely available. We are monitoring a few cases to try to achieve those goals, in particular with the recently launched SVOM X-rays satellite, which has also a visible telescope on board.

Because of the delay between the real start of the luminosity increase and the Gaia Alert, we cannot use the ascending phase to measure timescales, but try to do it on the descending phase, although the luminosity becoming fainter makes it a bit more difficult. Out of our large sample of CLQ's, we therefore follow a few specific cases to monitor further changes and try to measure some timescales. We show some illustrative examples below, where the light curve is steady, or starting slowly to decrease again. G19buq for instance (**Fig.5**), after the strengthening of the broad component since the earlier SDSS spectrum, does not show (yet?) further spectral changes although the light curve has started to show a decline. Another example is G20cvt: our spectra do not show any evolution of the broad lines in recent times, not even with respect to the archival spectrum mentioned by Motch et al. (1998), except for a possible small change in the [OIII] to Hbeta ratio. This indicates that the timescales for spectral evolution are of the order of a few years in most cases. We continue monitoring these objects and some more (e.g. 20bta, 20fku, etc…) to detect further evolution.

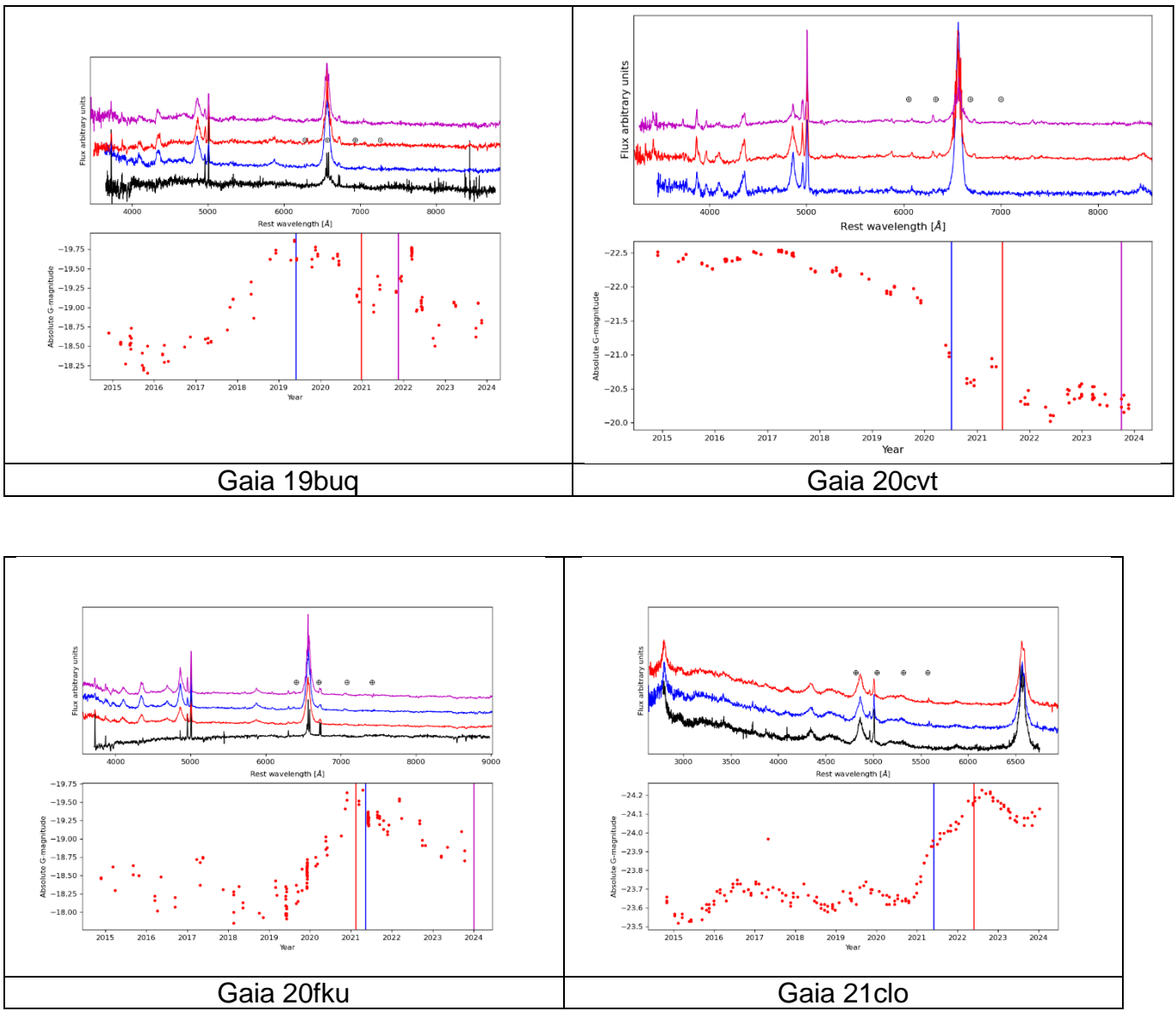

**Fig. 5** Monitoring of some new CLQ's

Same set-up as in Fig. 3d. The SDSS spectrum (if available) is in black, below our own spectra in color.

The colour of the spectrum corresponds to the color of the tick-mark in the light-curve.

## 3. A variety of light curves (flares, supernovae, moving dust blobs…) and discussion.

Besides the clear cases discussed above, such as TDE's or CLQ's, a diversity of light curves is observed in Gaia Alerts, as was the case in previous large samples. For example, in CRTS, Graham et al. (2017) showed examples of symmetric light curves, which they interpreted as single point microlensing events, although only 9 of their 51 events were satisfying their microlensing model. A number of Gaia light-curves of known AGN's appeared also to be quite symmetric, at least in the first times after their discovery: examples are Gaia 20cbv, or 20bta for instance **(Fig. 6a)**, but this initial symmetry disappeared later on. It shows the importance of long-term, photometric monitoring.

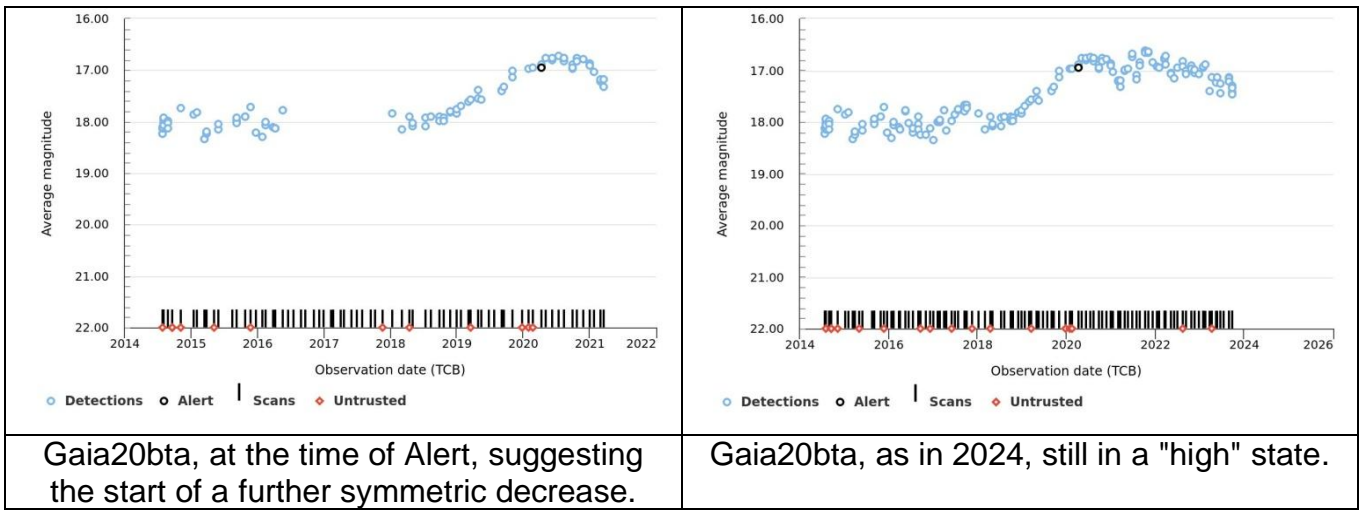

Fig. 6a  Evolution with time of the light-curve of Gaia 20bta. Gaia g magnitude versus time (in years); tick marks in abscissa indicate individual observations. The point in black is the Alert.

We found no example of a single, symmetric light curve interpretable as due to a microlensing event, except perhaps Gaia 22eps, which needs to be followed further.  There are some events showing a decreasing, symmetric light-curve, like Gaia 19aum **(Fig.6b):** this cannot be due to microlensing, but could be due to a temporary obscuration by transiting dust blobs. Its recent evolution might however suggest a sort of recurrence, or periodicity (possibly in a binary system?)

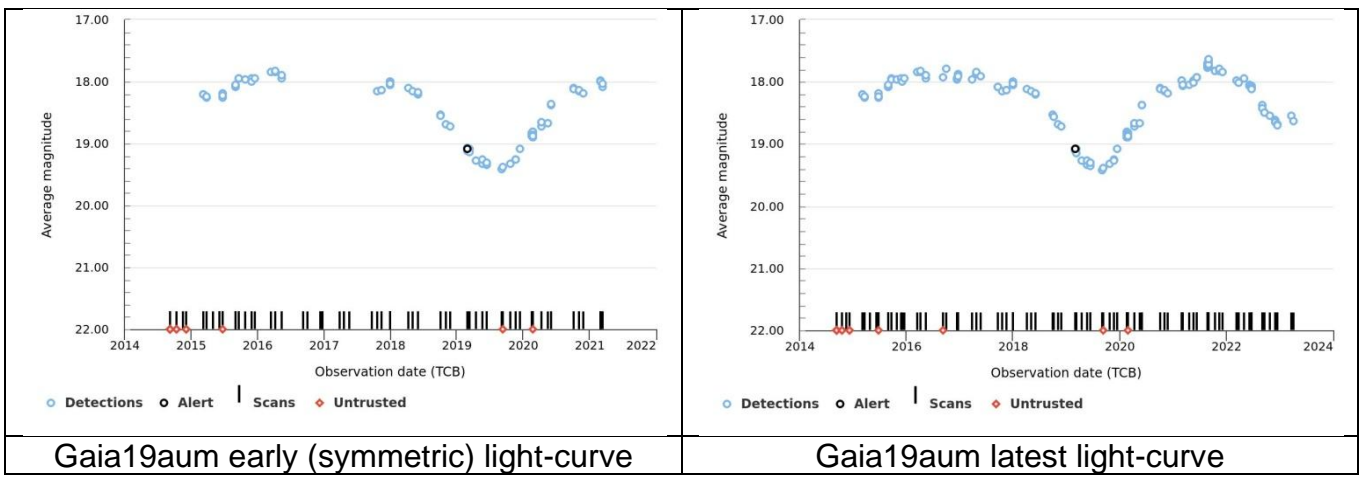

Fig. 6b  Evolution with time of the light-curve of Gaia 19aum. Same as in Fig. 6a

Other objects show asymmetric light curves, with a sharp rise followed by a slow decline, which could be due to a supernova, specially in Seyfert 2s which often also have starbursts episodes. The scarce, and irregular Gaia sampling does however not allow a SN classification from the light-curve alone, the episode lasting usually a few weeks only. Only when a spectrum could be obtained close enough to the maximum can the SN be identified, as was shown in two cases by Huo et al. (2020).  We note however that several cases of flares, where the classification spectrum, obtained long after the Alert, reveals a non-active galaxy, are fully compatible with a SN episode (see in particular the KR sample discussed in the Appendix).

In other cases, some narrow peaks appear, suggesting some short flaring is occurring. Examples of such are Gaia 20ccd, 20aen or 22cbn, but their light-curve is not reminiscent of any known type of transient. In the case of 18dmv, or 21cyz, these short events in quasars seem to repeat at some more or less regular intervals **(Fig. 7a)**, reminiscent of what is seen in some cataclysmic variables, possibly suggesting the existence of a variable gas reservoir close to the central black-hole. Some other objects show broader features in their light-curve, like 20dxo or 22bgh, where the temporary brightening could be due both to an increase of accretion and/or a diminishing of the obscuration, but the absence of simultaneous X-rays observations does not allow to conclude at this stage. Note that our successive spectra of 20dxo do not show any significant changes, but they were not immediately contemporaneous with the Alert. We could obtain, for some objects, an optical spectrum close to the Gaia Alert (for 21ahi, 21bhx, 21bij, 21bro, 21bww, 21bzs, 21eut and 21fug; see examples in **Fig. 7b**) but those spectra do not show any significant variation either. As many of these objects are also radio sources, some of these short-scale variations might be ascribed to fluctuations in a jet, a standard feature of BL Lac or similar objects and will be discussed in a companion paper.

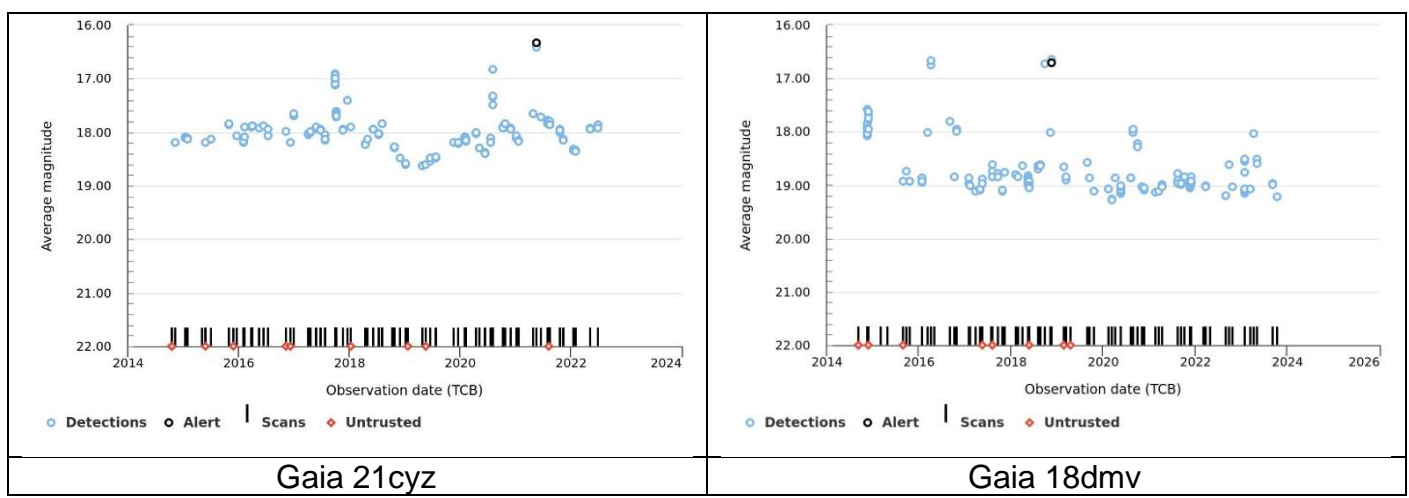

| Gaia 21cyz | Gaia 18dmv |

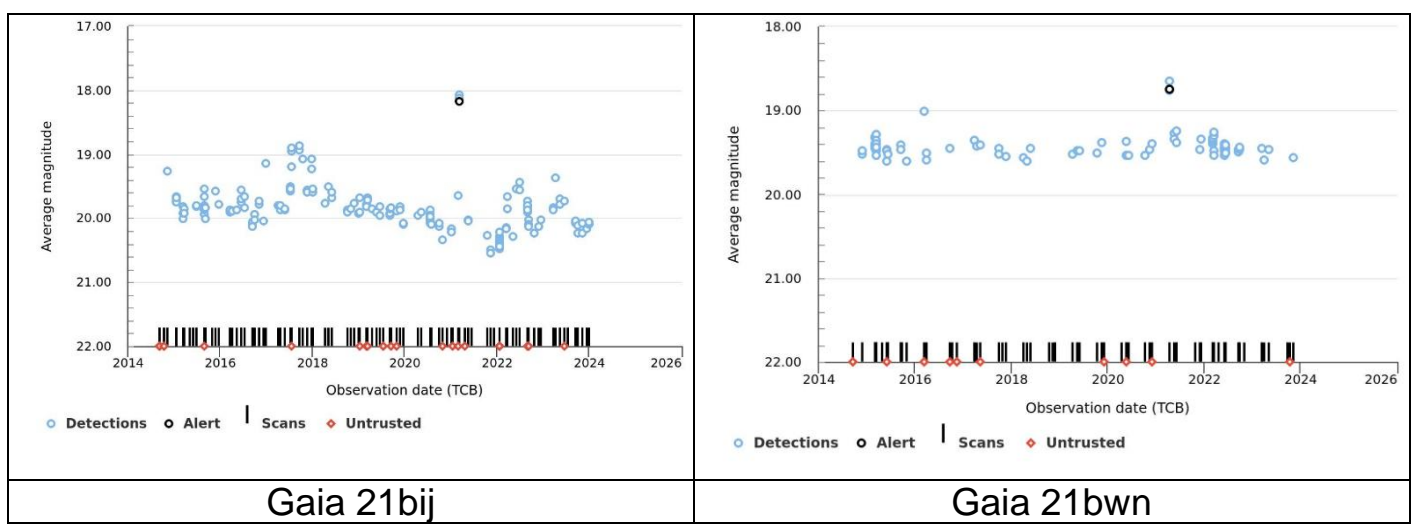

| Gaia 21bij | Gaia 21bwn |

> Fig.7 a,b  Examples of flaring events (same set-up as in Fig. 6)
>
> a (top): Gaia 21cyz and 18dmv   b (bottom): Gaia 21bij and 21bwn

Concerning CLQ's, McLeod et al. (2016) have argued, in their discussion (see previous section), that a single obscuring blob, drifting away from the line of sight, cannot be an explanation for a raising light curve, because, on one hand, the size of the central, continuum source is to large to be obscured by a single blob, and, on the other hand, even if that would be possible, the light curve should then have a flat top, which is never observed. Their use of a single blob is however a simplistic assumption, and there is some consensus today that the obscuring torus is, rather than a homogeneous and continuous assembly of obscuring matter, a mix of dust and gas whose distribution covers a range of sizes and different structures.  For instance, Esparza-Arredondo et al. (2021) discussed a set of well-studied, closeby AGN, with available optical, X-rays and IR data. They showed that a different mix of gas and dust structures is needed for each of them: the gas distribution is homogeneous when no variations are detected in the X-rays, but must be clumpy in the opposite situation. For the dust, their models often require a clumpy structure to best reproduce the observed data and the geometry of the torus is different from one object to the other. In such a frame, with multiple blobs transiting, progressive changes in obscuration could occur along the line of sight. To interpret the changes seen in our objects requires a multi-wavelength analysis, including data in the X-rays (to monitor column densities) and in the far-IR range (to monitor UV re-processed by dust), which are not yet available, but should come in the near future.

It is intriguing that we have so few CLQs in common with other surveys: only one with the CRTS sample of Graham et al. (2017, 2020), which identified 111 spectroscopically confirmed cases, although both their selection and ours were based on photometric variations. A selection bias is to be incriminated here: they principally looked for large flares, hence essentially bell-shaped light-curves, while Gaia Alerts were mainly based on continuously increasing (or decreasing) light curves. But we have no object in common either with the large sample (116 cases) of changing-look AGN of Zeltyn et al. (2024), who selected their candidates not on photometric but on spectroscopic changes within successive SDSS spectra: the situation may change in the future, as they looked only at the first year of SDSS-V repeated spectroscopy.  And we have only one in common with the sample of 61 of Green et al.  (2022), also selected on spectroscopic changes only: but they looked preferentially at broad line quasars, hence selected dimming objects, while our follow-up selected preferentially objects with increasing brightness. As spectral dates, and photometric alerts dates (Gaia or other surveys) are quite often distant in time, it makes the interpretation more difficult. The differences between those various samples show that selection effects are important, and are an indication that different phenomena are possibly at work in the centers of AGN. Time scales of changes, and time differences between photometric and spectral changes need to be measured and monitored accurately.

# Conclusions

This study of over 8 years of Gaia Alerts in galaxy nuclei confirms that variability is a powerful tool to detect both new AGN and spectacular changes in spectral properties of already known ones. Although only about 10% of the candidates could be followed spectroscopically from the ground, we have confirmed 64 new quasars or AGN. We detected also 56 new, secure cases of Changing Look in known quasars (including those from the KR sample, see Appendix), plus 23 more which need to be confirmed with better spectra. Several dozen more are monitored to measure quantitatively possible changes. This shows that the phenomenon of "changing look" is more common than initially thought.  In view of the incomplete Gaia selection effect, coupled with a variable human vetting process, it is impossible to provide accurate statistics of the occurrence of this phenomenon. If we assume that our partial selection for follow-up is representative of the whole population of variable quasars, and that the Gaia selection is itself representative, we come up with a proportion of approximately 10% of the variable quasars which present the changing look phenomenon. As various types of variations are detected (beyond the obvious cases of CLQ's), several physical phenomena must be occurring:  various changes in the accretion rate (short flares, TDE's, winds, etc...), changing obscuration or modifications of the BLR structure, SNe,... As the various data (Gaia alert, optical follow-up, X-rays or IR observations) were not contemporaneous, it is difficult to identify the process at play. A systematic, multi-wavelength follow-up of some representative cases, from the X-rays to the IR, is necessary to distinguish the various possibilities. Some objects are already followed-up with existing facilities to identify further changes. This will be even easier in the near future with the recently commissioned Vera Rubin telescope, together with space observations from the new SVOM gamma- and X-rays satellite, and the expected Nancy Roman IR space telescope.

Acknowledgments

We thank Cl. Pellouin (IAP) and M. Ward (Durham) for useful exchanges. This paper made extensive use of the Gaia Alerts, developed by the Photometric Science Alerts Team (http://gsaweb.ast.cam.ac.uk/alerts) with data from the ESA Gaia satellite and DPAC. It also used the NASA/IPAC  Extragalactic  Database (NED), which is funded by NASA and operated by the California Institute of Technology. Reference spectra mostly come from the Sloan Digital Sky Survey, funding of which has been provided by the Alfred P. Sloan Foundation, the U.S. Department of Energy Office of Science, and the Participating Institutions. SDSS acknowledges support and resources from the Center for High-Performance Computing at the University of Utah. The SDSS web site is www.sdss4.org. Based on observations made with the Nordic Optical 2.5m Telescope, owned in collaboration by the University of Turku and Aarhus University, hosted at the IAC in Spain, and with the Haute-Provence 1.93m telescope, owned by CNRS and hosted by OSU-Pytheas in south of France.

Spectra will be available on TNS and Github

## Appendix:  the KR sample

Kostrzewa-Rutkowska et al. (2018) (hereafter KR) published a list of 482 Gaia Nuclear Transients (GNT) selected for variations over a one year period with a different algorithm, shown to be more efficient in center of galaxies than the one used for standard Gaia Alerts (see discussion in the KR paper). Among them over 350 were labelled "unknown" and the remaining ones are galaxies or QSO's as indicated from their SDSS spectra. We started to observe some of the "unknown" ones in order to identify their nature, essentially during 2018 and 2019, with the same setting as for the Gaia alerted objects and obtained 73 spectra. Since the publication of the KR paper in 2018, and while our observations were on-going, some of their objects received an SDSS classification in later installments (serving as reference for further variations) so that only 34 remained "unknown" among the 73 we observed. Among those 34, we identify "17 QSO's plus 5 possible ones (poor S/N or single line only), all presented in **Table 1**, with examples in **Fig. 8a**. The 12 others show standard galaxies spectra: it is possible that some of those were alerted because of a SN, but as our spectra are distant in time from the Alert, we cannot conclude about the nature of that Alert. The scarce sampling by Gaia does not allow to conclude from the light-curve alone whether it was indeed a SN or not, specially because some of the confirmed quasars have quite similar light-curves, with a short, narrow (one or two points only) excursion above the average magnitude which could also be a flare in the AGN. Among the 27 candidates with light-curves of the "flaring" type we examined, 11 proved at the end to be a quasar and are also in Table 1. By contrast, some objects marked as "galaxy" (or some "unknown") do indeed show a galaxy-type spectrum (without AGN features) in our follow-up, and could thus have been alerted because of a SN event. As the available light-curve of the KR objects covers a one year period only, and was not monitored further by Gaia Alerts, it is not possible either to identify a genuine AGN from its long term behavior alone.

.   For those which have a reference SDSS spectrum, our second epoch spectrum allows us to identify 11 new CLQ's (also presented in **Table 2**), with definite changes in the BLR, the NLR and/or the blue continuum and their properties do not seem to be different from those described above in the main Gaia Alert scheme. Examples are shown in **Fig. 8b**. Ten others show possible minor changes only, and need to be measured more accurately. As we followed only a fraction of the KR sample, it is difficult to extract good statistics from it, but the numbers given above are not markedly different from the proportions we derive from the main Gaia Alert stream. But in view of the total numbers in the KR sample, as compared to the Gaia Alerts during the same period, it is clear that this new method seems to be more efficient to detect nuclear transients but, regrettably, could not be implemented further.

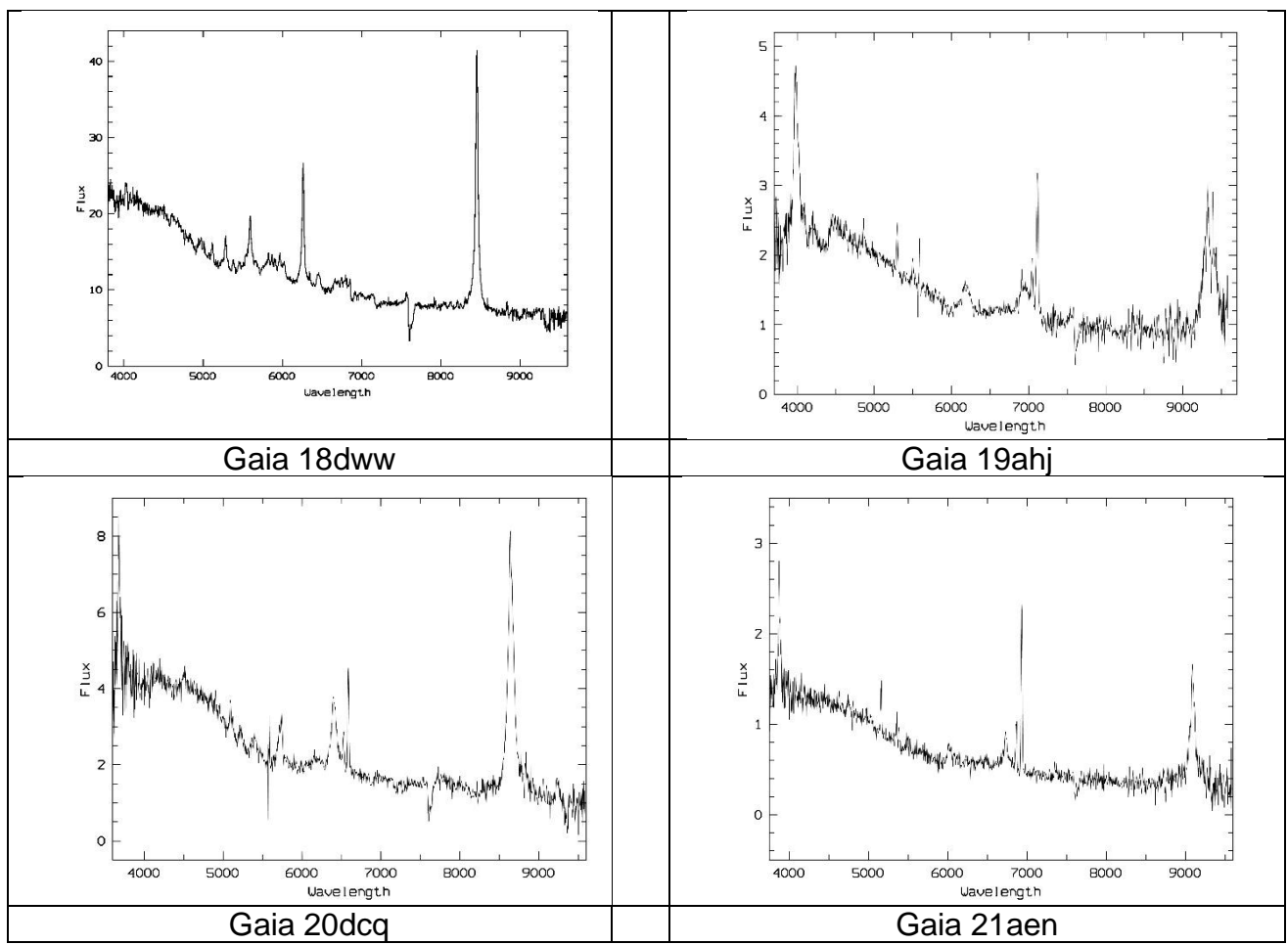

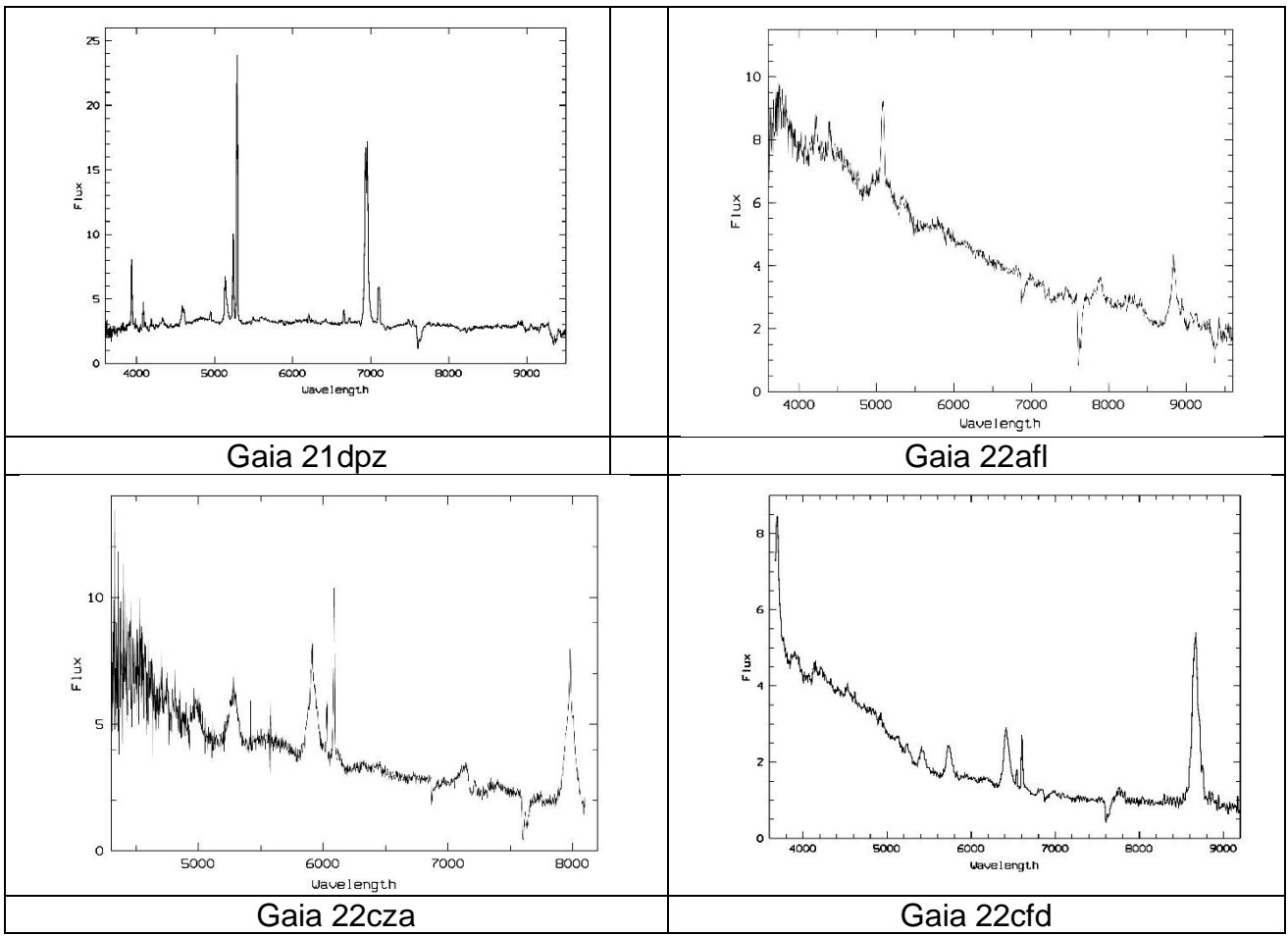

Fig. 1 Examples of classification spectra. Flux is in ordinates (in units of 10E-16 ergs/cm$^2$/s/Å), versus wavelength (in Å) in abscissa.

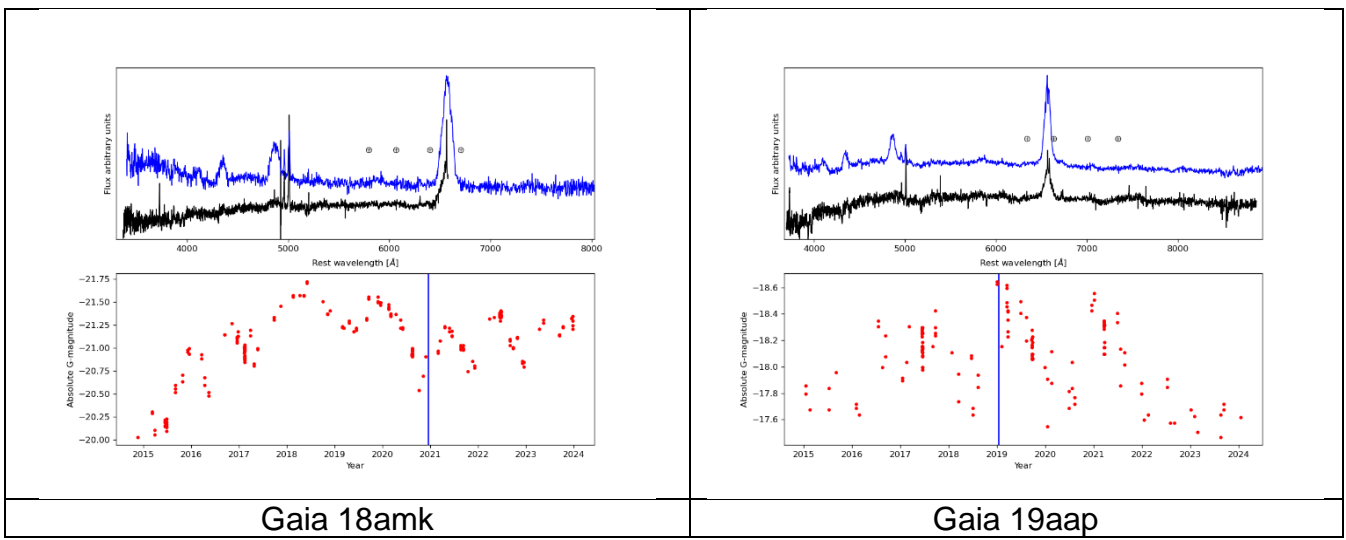

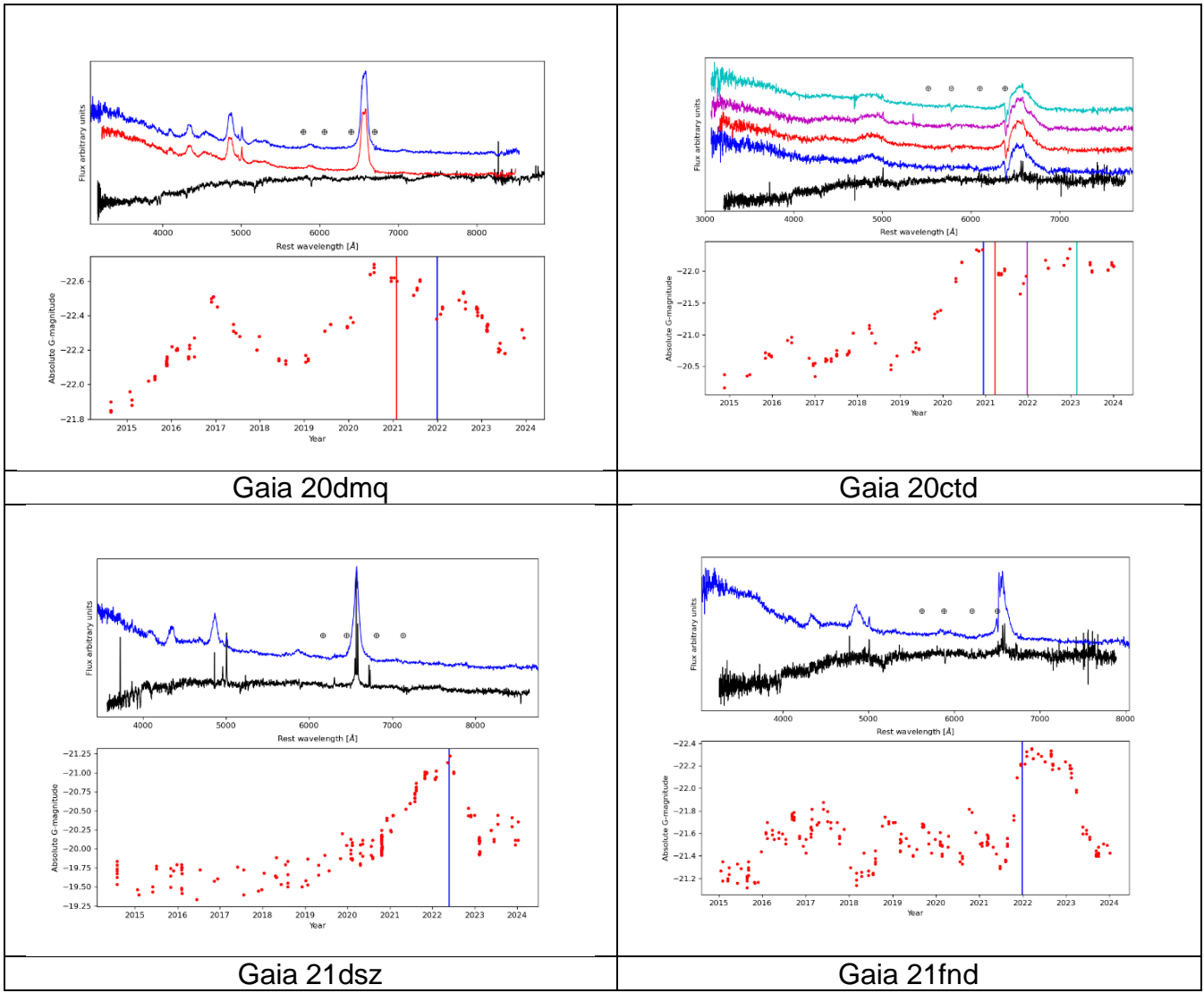

Fig.4  Examples of CLQ's (same as in Fig.3d)

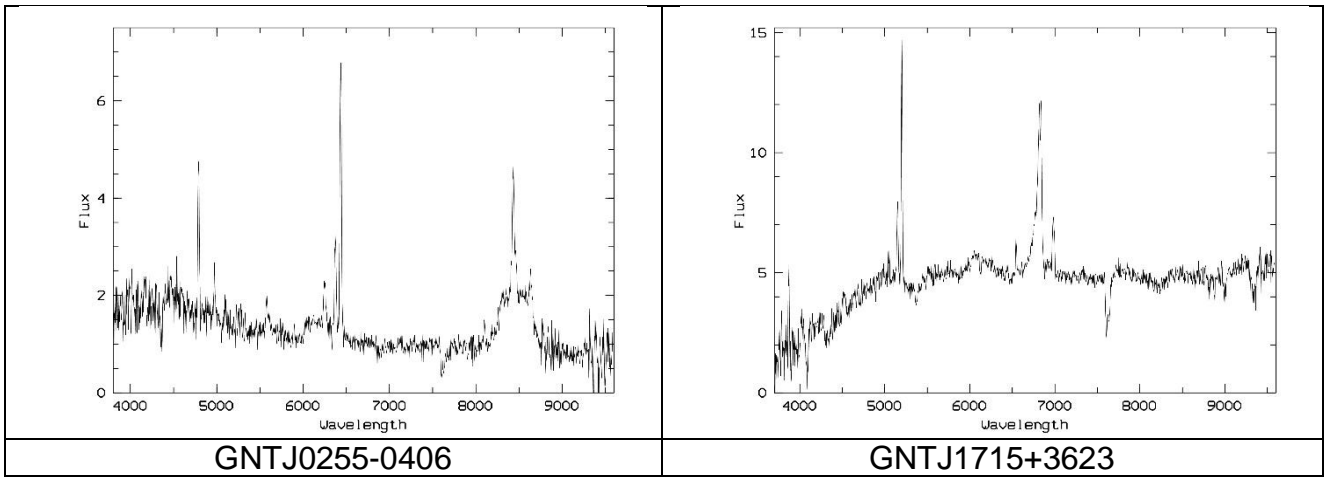

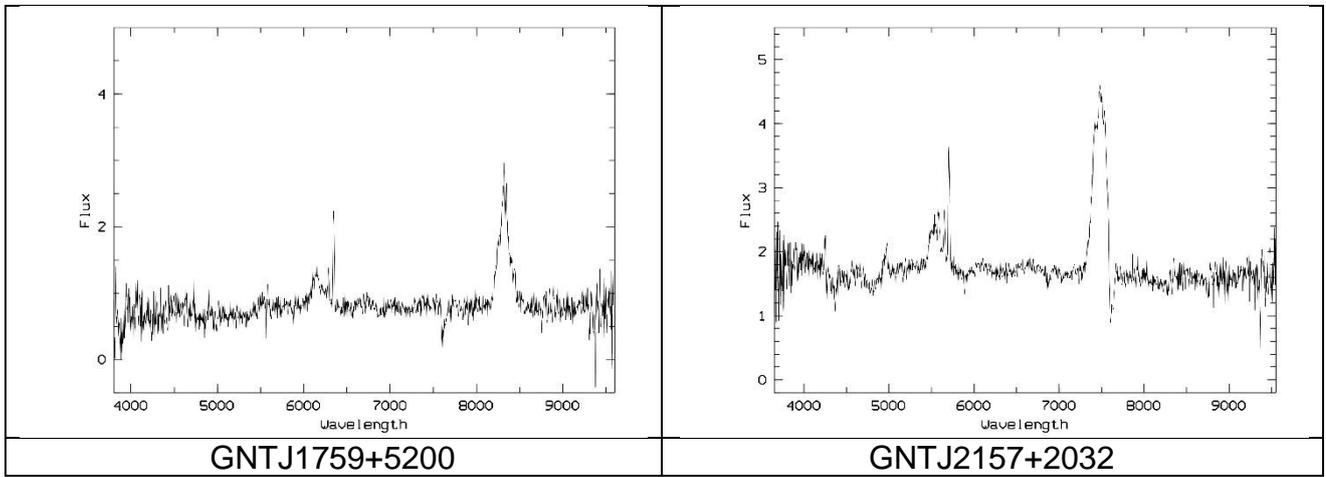

Fig. 8a: Examples of classifications from the KR sample (same as Fig. 1)

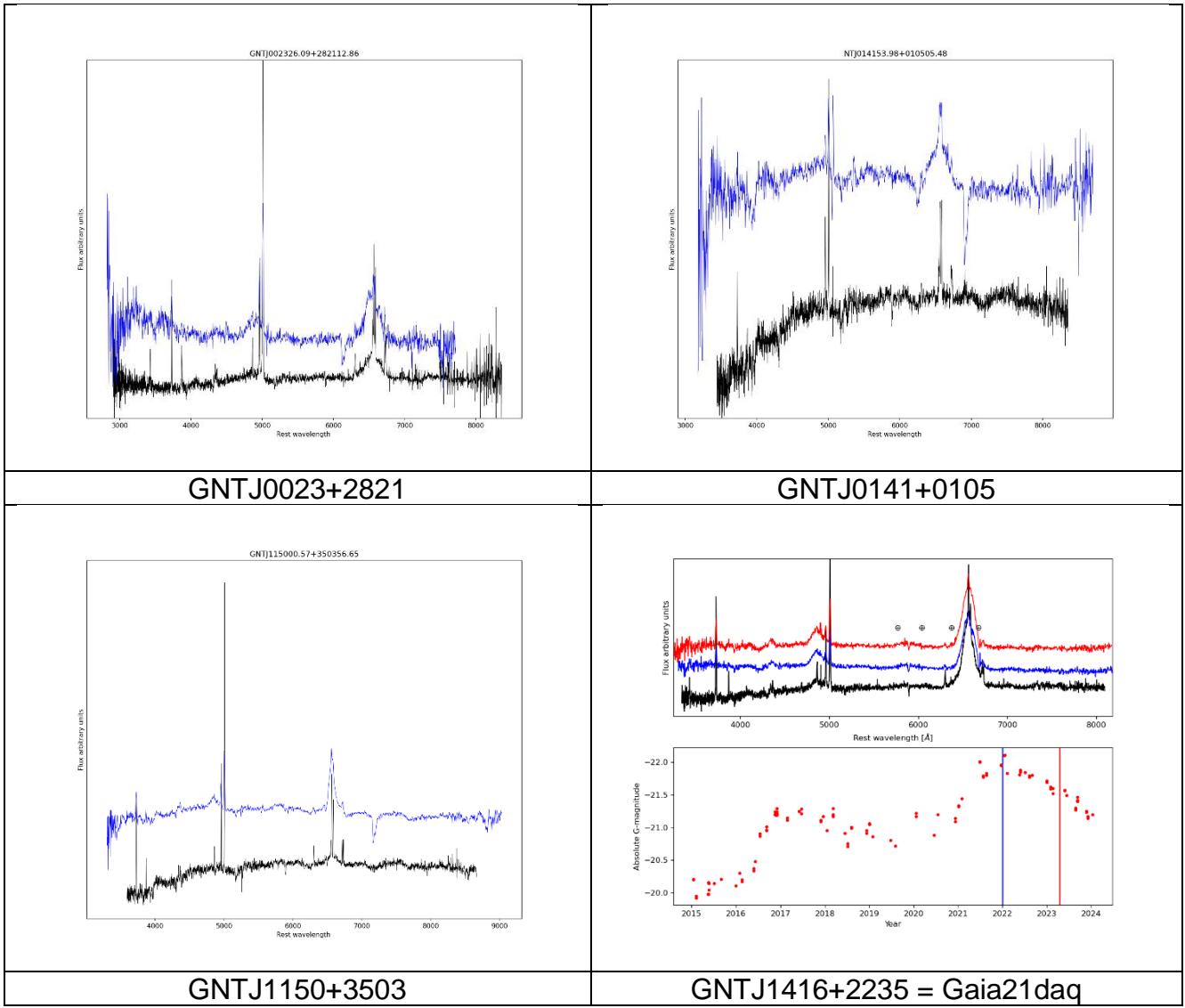

Fig. 8b: Some CLQ's among GNT alerts in the KR sample (same as Fig. 4)

Ethics declaration: Not applicable

| Gaia Name | IAU name | Alpha | Delta | Alert date | Spec. Date | Gaia class | Our Classification | z | Lines seen | z_Quaia | Radio/optical | Comments |
|---|---|---|---|---|---|---|---|---|---|---|---|---|
| 15aai | — | 233.55516 | 62.98277 | 22/1/2015 | 29/6/2022 | AGN | QSO | 0.237 | [OIII], Hb, Hg | 0.4074* | -- | |
| 16aoz | 2016eaf | 162.11514 | 71.72666 | 15/5/2016 | 20/04/2020 + | 7C QSO | QSO | 1.141 | | 0.4521* | 375 | |
| 16bpi | 2016anj | 107.17241 | -24.27922 | 21/10/2016 | 21/12/2020 | cand. AGN | Star | | | -- | -- | Star |
| 17cff | 2017fro | 259.98272 | 41.68041 | 3/9/2017 | 27/4/2023 | Nuclear Flare | NLS1 | 0.163 | [SII], Ha, HeI, FeII, [OIII], Hb, FeII, Hg, [NeIII],[OII] | -- | -- | |
| 17dgc | 2017iwk | 113.97017 | -51.12044 | 12/9/2017 | — | cand. AGN | Sey1 | 0.149 | Ha, [OIII], Hb, Hg | 0.1597 | -- | From 6dF |

**Table 1** Spectroscopically confirmed AGN and quasar candidates, alerted by Gaia because of variability (first lines only, full table in electronic form). Columns: 1: Gaia name; 2: IAU name; 3-4: J2000 coordinates in degrees; 5: Alert date; 6: Date of our spectrum (a + sign indicates more than one date available); 7: proposed Gaia classification; 8: our classification; 9: measured redshift; 10: lines used for the redshift determination; 11: z from Quaia (empty if not available; an asterisk indicates a discrepant value with our own measurement); 12: ratio of radio (1.4 GHz) to optical V flux, a – sign indicates a ratio less than 10 percent; 13: Comments

| Gaia Name | IAU Name | Alert Date | Spectral date | SDSS Date | Gaia Class | Broad Hα | [NII]/Hα | Broad Hβ | [OIII]/Hβ | Blue Continuum | Mg II |
|---|---|---|---|---|---|---|---|---|---|---|---|
| 1 | 2 | 3 | 4 | 5 | 6 | 7 | 8 | 9 | 10 | 11 | 12 |
| 16bwn | 16ihw | 19/11/2016 | 5/9/2020 | 24/9/2011 | QSO | No significant change | | | | | |
| 18amk | 18xu | 17/2/2018 | 16/12/2020 | 5/12/2007 | Sey 1 | Increase | Decrease | Incr | Decr | small incr | |
| 18cdf | 18fat | 11/8/2018 | 26/8/2018 | 12/6/2010 | QSO | No significant change | | | | | |
| 18dbv | 18hmb | 18/10/2018 | 6/11/2018 | 27/2/2004 | QSO | No significant change | | | | | |
| 18dct | 18hmu | 20/10/2018 | 6/11/2018 | 18/12/2004 | QSO | | | small decr. ? | small decr. ? | small decr. ? | |
| 18ded | 18hpc | 26/10/2018 | 6/11/2018 | 13/6/2015 | QSO | No significant change | | | | | |
| 18dsi | 18jni | 6/12/2018 | 2/1/2019 | 28/2/2006 | QSO | | | | | small increase | small incr.? |

**Table 2** Variations seen in our spectra with respect to an earlier reference spectrum (first lines only, full table in electronic form). Visual estimates (small means less than 10 percent change). Columns: 1: Gaia name; 2: IAU name; 3: Alert date; 4: Date of our spectrum; 5: Date of SDSS spectrum; 6: proposed Gaia classification; 7-12: estimates of the variations of the broad component of Hα, [NII]/Hα ratio, broad Hβ component, [OIII]/Hβ ratio, intensity of the blue continuum, and of the MgII line; 13: Comments.